\documentclass[namedreferences]{solarphysics}
\usepackage[hyperref,optionalrh]{spr-sola-addons} 
\usepackage{graphicx} 
\usepackage{color}   
\ifx \doiurl    \undefined \def \doiurl#1{\href{http://dx.doi.org/#1}{\textsf{DOI}}}\fi
\ifx \arxivurl  \undefined \def \arxivurl#1{\href{http://arxiv.org/abs/#1}{\textsf{arXiv: #1}}}\fi
\ifx\urlurl    \undefined \def\urlurl#1{\href{http://#1}{\textsf{#1}}}\fi

\begin{document}
\begin{opening}
\title{Linkage of Geoeffective Stealth CMEs Associated with the Eruption of Coronal Plasma channel and Jet-Like Structure}
\author[addressref=aff1,corref,email={sudheerkm.rs.phy16@itbhu.ac.in}]{\inits{S.K.}\fnm{Sudheer K.}~\lnm{Mishra}\orcid{https://orcid.org/0000-0003-2129-5728}}
\author[addressref=aff1]{\inits{A.}\fnm{A.K.}~\lnm{Srivastava}}
\address[id=aff1]{Department of Physics, Indian Institute of Technology (BHU), Varanasi-221005, India.}
\runningauthor{S. Mishra and A.K. Srivastava}
\runningtitle{Geoeffective Stealth CME}
\begin{abstract}
				We analyze the eruption of a coronal plasma channel (CPC) and an overlying flux rope using \textit{Atmospheric Imaging Assembly/Solar Dynamic Observatory} (AIA/SDO) and \textit{Solar TErrestrial RElations Observatory} (STEREO)-A spacecraft data. The CPC erupted first with its low and very faint coronal signature. Later, above the CPC, a diffuse and thin flux rope also developed and erupted. The spreading CPC further triggered a rotating jet-like structure from the coronal hole lying to its northward end. This jet-like eruption may have evolved due to the interaction between spreading CPC and the open field lines of the coronal hole lying towards its northward foot-point. The CPC connected two small trans-equatorial coronal holes lying respectively in the northern and southern hemisphere on either side of the Equator. These eruptions were  collectively associated with the stealth-type CMEs and CME associated with a jet-like eruption. The source region of the stealth CMEs lay between two coronal holes connected by a coronal plasma channel. Another CME was also associated with a jet-like eruption that occurred from the coronal hole in the northern hemisphere. These CMEs evolved without any low coronal signature and yet were responsible for the third strongest geomagnetic storm of Solar cycle 24. These stealth CMEs further merged and collectively passed through the interplanetary space. The compound CME further produced an intense geomagnetic storm (GMS) with Dst index= -176 nT. The $z$-component of the interplanetary magnetic field [$B_{z}$] switched to negative (-18 nT) during this interaction, and simultaneous measurement of the disturbance in the Earth's magnetic field (Kp=7) indicates the onset of the geomagnetic storm.
\end{abstract}
\end{opening}
\section{Introduction}
The study of coronal mass ejections (CME) and their association with solar prominences and flares is one of the most important current problems in solar physics. The internal dynamics and MHD instability play a significant role in the eruption of solar prominences (\textit{e.g.} Srivastava {\it et al.}, 2010; Joshi {\it et al.}, 2013; Mishra {\it et al.}, 2018; Mishra and Srivastava, 2019). A causal relationship between solar prominences and CMEs has been observed (\textit{e.g.} Gopalswamy {\it et al.}, 2003; Filippov and Koutchmy, 2007; Schmieder, D\'emoulin, and Aulanier, 2013; Webb, 2015). Various observational and theoretical models have been used to understand the triggering mechanism of coronal mass ejections. Some of the most recognized models for CME triggering are ``tether cutting or flux cancellation mechanism'' (\textit{e.g.} Moore and LaBonte, 1980; van Ballegooijen and Martens, 1989; Moore {\it et al.}, 2001; Amari {\it et al.}, 2003), ``shear motions'' (\textit{e.g.} Barnes and Sturrock, 1972; Low, 1977; Aly, 1990; Deng {\it et al.}, 2001; Kusano {\it et al.}, 2004), ``the magnetic breakout model'' (\textit{e.g.} Antiochos, DeVore and Klimchuk, 1999; Lynch {\it et al.}, 2004; Sterling and Moore, 2004;  Archontis and Torok, 2008) ``emerging flux and injection mechanism'' (\textit{e.g.} Chen, 1989; Chen and Shibata, 2000), and ``hybrid mechanism'' (Amari {\it et al.}, 2000). There are some observational features, by which we identify the coronal mass ejections (CMEs). These observational features of CMEs are morphology and mass, angular width, occurrence rate, velocity and energy, and association with the filament and flares. Chen (2011) and Webb and Howard (2012) have reviewed the observational and theoretical view of CMEs. The observational view of CMEs in the heliosphere has been discussed using remote-sensing coronagraphic data (Gopalswamy, 2004). Recently, the origin, predictability, and the evolution of solar eruptions have been discussed in great detail (Green {\it et al.}, 2018). Based on the morphological evolution, the CMEs are classified as halo CMEs, partially halo CMEs, narrow CMEs, and CMEs with low coronal signatures. A typical CME consists of three components; an eruptive prominence associated with a bright core, a lower density dark cavity region, and a diffuse leading-edge CME front with its legs connected to the Sun. Robbrecht, Patsourakos, and Vourlidas, (2009) use \textit{Solar TErrestrial RElations Observatory} (STEREO) data to discuss the evolution of a CME that has no signature in the lower corona. Because there is no signature present in the low corona, it is difficult to find their eruptive source on the solar disk. Therefore, it makes it difficult to forecast space weather, without any initial signature of the eruption in the lower corona. Because there is no signature in the inner corona, this type of CMEs are known as stealth CMEs.   \\

The stealth CMEs typically propagate with a speed less than 500 km s$^{-1}$ and are originated from the quiet-Sun region (\textit{e.g.} Ma {\it et al.}, 2010; D'Huys {\it et al.}, 2014). The recent study concludes that the source region of the stealth CME is associated with the active region of the Sun also (O'Kane {\it et al.}, 2019). The source region of the stealth CME may also be located near the coronal holes (\textit{e.g.} Nieves-Chinchilla {\it et al.}, 2013; Lynch {\it et al.}, 2016; Nitta and Mulligan, 2017). Bhatnagar (1996) has discussed the solar eruption from the coronal holes. He suggested that the open field lines of the coronal hole may reconnect with the overlying filament and this triggers the solar eruption. Solar filaments disappears in the trans-equatorial coronal holes (Chertok {\it et al.}, 2002). The coronal holes act as a magnetic wall that resists the CME propagation (Gopalswamy {\it et al.}, 2009). An eruption arises from the coronal hole, which is associated with the mini-CME (\textit{e.g.} Innes, McIntosh, and Pietarila, 2010; Adams {\it et al.}, 2014; Moore, Sterling, and Falconer, 2015; Sterling {\it et al.}, 2015). Magnetic reconnection may occur near coronal holes (Yang {\it et al.}, 2011; Adams {\it et al.}, 2014) and may cause the eruption of the CMEs. They observe the impact of stealth CMEs at 1 {\,AU}. The rotational motions may be associated with the magnetic untwisting of solar jets (Moore, Sterling, and Falconer, 2015), which may trigger rotating jet-like CMEs mimicking the signature of stealth CME also. The first observational evidence of the rotating CME into the outer corona has been observed, which is related with the rotating jet-like eruption (\textit{e.g.} Yurchyshyn, Abramenko, and Tripathi, 2009; Vourlidas {\it et al.}, 2011; Joshi {\it et al.}, 2018). Stealth CMEs may have their in the rotating plasma structures in the inner solar corona or the evolution of faint flux-rope eruption. \\

CME observations are useful to study the Sun--Earth relation (Gosling, 1993; Harrison, 1995). The major difficulties in space-weather forecasting are to predict the intensity of a geomagnetic storm on the basis of the given solar input such as the source region, location, velocity, and types of CMEs. Several studies have been performed to observe the association of CME properties and the geoeffectiveness. Srivastava and Venkatakrishnan, (2002) studied the relationship between CME speed and intensity of the geomagnetic storm. They found that the initial speed of the CMEs is well correlated with the Disturbance storm time (Dst) index intensity, which is associated with the geomagnetic storm. Some solar eruptions occur without any signs on the disk but may lead to unpredictable geomagnetic activity near the Earth. The early warning for space weather activities does not arise in such types of events. Ma {\it et al.} (2010) have statistically analyzed the source location of the CMEs during solar minima and find that one-third of the CMEs are stealth type. In the stealth CMEs, approximately half of them are blowout type CMEs. The observational signature (appearance in coronagraphs, position angle, velocity profile, angular width, and scale invariance) of the stealth CMEs has been discussed by D'Huys {\it et al.} (2014). Pevtsov, Panasenco, and Martin (2011) observed that the filament channel without filament was the stealth CMEs source region. These CMEs are modeled and compared with \textit{Large Angle and Spectrometric Coronagraph/Solar and Heliospheric Observatory} (SOHO/LASCO) and \textit{Solar Terrestial Relations Observatory/Sun Earth Connection Coronal and Heliospheric Investigation} (STEREO/SECCHI) coronagraphic imaging data (Lynch {\it et al.}, 2016). The source region of the stealth CMEs may be located near open field lines or coronal holes region (Lynch {\it et al.}, 2016; Nitta and Mulligan, 2017). A new image-processing technique with high cadence has been used to study the onset and temporal evolution of stealth CME with low coronal signatures (Alzate and Morgan, 2017). Recent studies suggest that stealth CMEs may erupt from the active regions also (O'Kane {\it et al.}, 2019). Zhang {\it et al.} (2007) have discussed the connection between the solar eruption and interplanetary sources of an intense geomagnetic storm (Dst\textless-100 nT). They found that 12\,\% of total CMEs are launched without low coronal signature in Solar Cycle 23. The Earth-affecting CMEs that have a low-coronal signature rise with slow velocity (\textit{e.g.} Robbrecht, Patsourakos, and Vourlidas, 2009; Ma {\it et al.}, 2010; D'Huys {\it et al.}, 2014; Kilpua, Koskinen, and Pulkkinen, 2017; Lynch {\it et al.}, 2016; Nitta and Mulligan, 2017).\\

The present work, to the best of our knowledge, is the first extensive effort to understand the development of stealth CMEs, identification of their source regions, \textit{i.e.} its association with the coronal plasma channel (CPC) and a rotating jet-like eruption in the inner corona. Initially, a CPC-associated CME has appeared in the outer corona without any signature in the lower corona; however, the spreading CPC and post-eruption arcade indicate that an eruption has occurred from this region. This eruption further triggers flux rope associated CME. Later, a rotating jet-like eruption-associated CME from the coronal hole lying at the northward end of the CPC first appeared in the STEREO-A/COR2 FOV on 20 August 2018, again without any signature in the lower corona. These three CMEs further interact with each other and form a bigger compound CME in the outer corona. It passes through the interplanetary space and interacts with the Earth magnetosphere. The observed CME is a stealth type, which is very diffuse and evolves with slow velocity. However, it is responsible for the onset of the third most intense geomagnetic storms in the Solar Cycle 24. In this article, we propose that an intense geomagnetic storm may have occurred due to the stealth CMEs, which are associated with a hot coronal plasma channel, a flux rope, and a rotating jet-like eruption from the coronal hole. The observations and data analysis are discussed in Section 2. Section 3 is framed with the observational results corresponding to the development of filament over the coronal hole, coronal plasma channel (CPC) associated eruption, the flux rope and rotating jet-like eruptions, the formation of stealth CMEs, and propagation of the compound CME in the interplanetary space, and lastly its impact on Earth's magnetosphere to create geomagnetic storm. In the last section, discussion and conclusions are presented.
 \begin{figure*}
\hspace{-1.0cm}
\includegraphics[scale=1.0,angle=0,width=13.0cm,height=13.0cm,keepaspectratio]{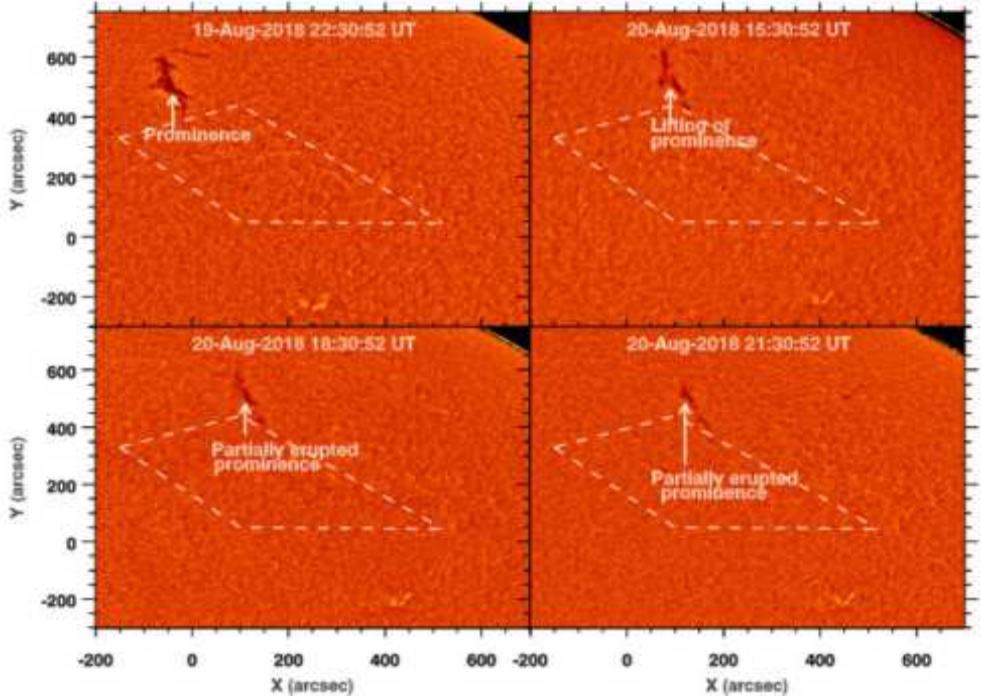}
\caption{Evolution of a quiescent filament as observed by GONG/H$\alpha$. It evolved spatially and temporally and crossed the central meridian of the Sun on 26 August 2018. The filament was lifted and partially erupted when it passes over the northward lying coronal hole. The coronal plasma channel (CPC) is associated with the filament in its southwest direction, and its location is shown in the {\it white box}. It appears to be empty as no traces of cool plasma in H${\alpha}$ has been found in CPC, and it is only visible in the coronal emissions.}
\end{figure*}
\section{Data and Analysis}
We used \textit{Global Oscillation Network Group} (GONG: Harvey {\it et al. 2011}) H$\alpha$ data from Big Bear Solar Observatory (BBSO) site. The Big Bear Solar Observatory (BBSO) provides full-disk imaging data of one minute cadence and 2$''$ spatial resolution (Denker {\it et al.}, 1999). We download the GONG H$\alpha$ data from \urlurl{sdac.virtualsolar.org/cgi-bin/vsoui.pl} to observe the evolution of the filament (Figure 1). \\
We used high-resolution data of \textit{Atmospheric Imaging Assembly} (AIA: Lemen {\it et al.}, 2010) onboard the \textit{Solar Dynamic Observatory} (SDO: Pesnell, Thompson, and Chamberlin, 2012). The AIA contains seven extreme ultraviolet, two ultraviolet, and a visible region full-disk solar imaging data with a 1.5$''$ spatial resolution and 12 second temporal resolution. We used 171 {\AA}, 193 {\AA}, 211 {\AA}, and 304 {\AA} EUV imaging data for our analysis. We have selected 16-hour temporal data on 20 August 2018 to observe the dynamics of the coronal plasma channel (CPC), associated faint flux rope, and a jet-like eruption.\\
\begin{figure*}
\hspace{-1.0cm}
\includegraphics[scale=1.0,angle=0,width=13.0cm,height=13.0cm,keepaspectratio]{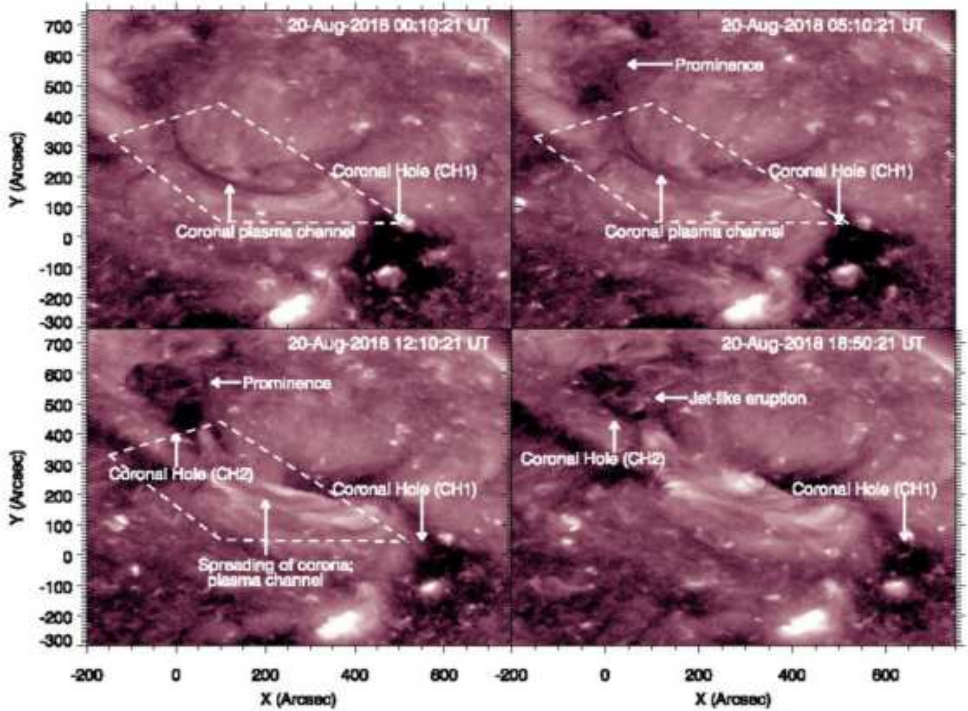}
				\caption{The full field of view (FOV) of the coronal plasma channel (CPC) that connects the two coronal holes (CH1 and CH2) observed by SDO/AIA 211{\AA} on 20 August 2018. The spreading coronal plasma channel {\it (white box)} and post-eruption arcade indicate that an eruption from this region has already occurred. The spreading coronal plasma channel further interacts with an open field line of a northward lying coronal hole (CH1) and may trigger a jet-like eruption {\it (lower panel)}. The {\it white-box-region} consists of a coronal plasma channel (CPC). The impact of an eruptive coronal channel could be observed in the form of spreading of this plasma channel and formation of the post-eruption arcade as seen within the \textit{white box region}.}
\end{figure*}
We used \textit{Sun Earth Connection Coronal and Heliospheric Investigation} (SECCHI) data from instruments onboard the \textit{Solar Terrestrial Relations Observatory} (STEREO-A) spacecraft to observe the inner corona up to the interplanetary space. The \textit{Extreme Ultraviolet Imager} (EUVI: Wuelser {\it et al.}, 2004) observe the lower corona (1\,--\,1.7\,R$_\odot$) with five-minute temporal cadence and 1.6$''$ spatial resolution. STEREO-A/COR2 (Howard {\it et al.}, 2008) data has been used to observe the outer corona from
2.5\,--\,15\,R$_\odot$ with 14.7$''$ spatial resolution and 30-minute temporal resolution. The Coronal plasma channel (CPC) associated CME, the CME associated with the faint flux rope and rotating jet-like structure have also been evident into the outer corona as observed by STEREO-A/COR2. Interplanetary space has been observed by STEREO-A/HI1 and HI2 (Howard {\it et al.}, 2008). HI-1 observes in visible light from the outer corona and up to the lower interplanetary space (15\,--\,90\,R$_\odot$). The observed multiple CMEs are merged within each other in this region and propagate towards the interplanetary space in the form of a compound CME. The interplanetary space has been observed by HI2 from 90\,R$_\odot$\,--\,4\,AU with two hours of temporal cadence.\\
\begin{figure*}
\includegraphics[scale=1.0,angle=0,width=18.0cm,height=18.0cm,keepaspectratio]{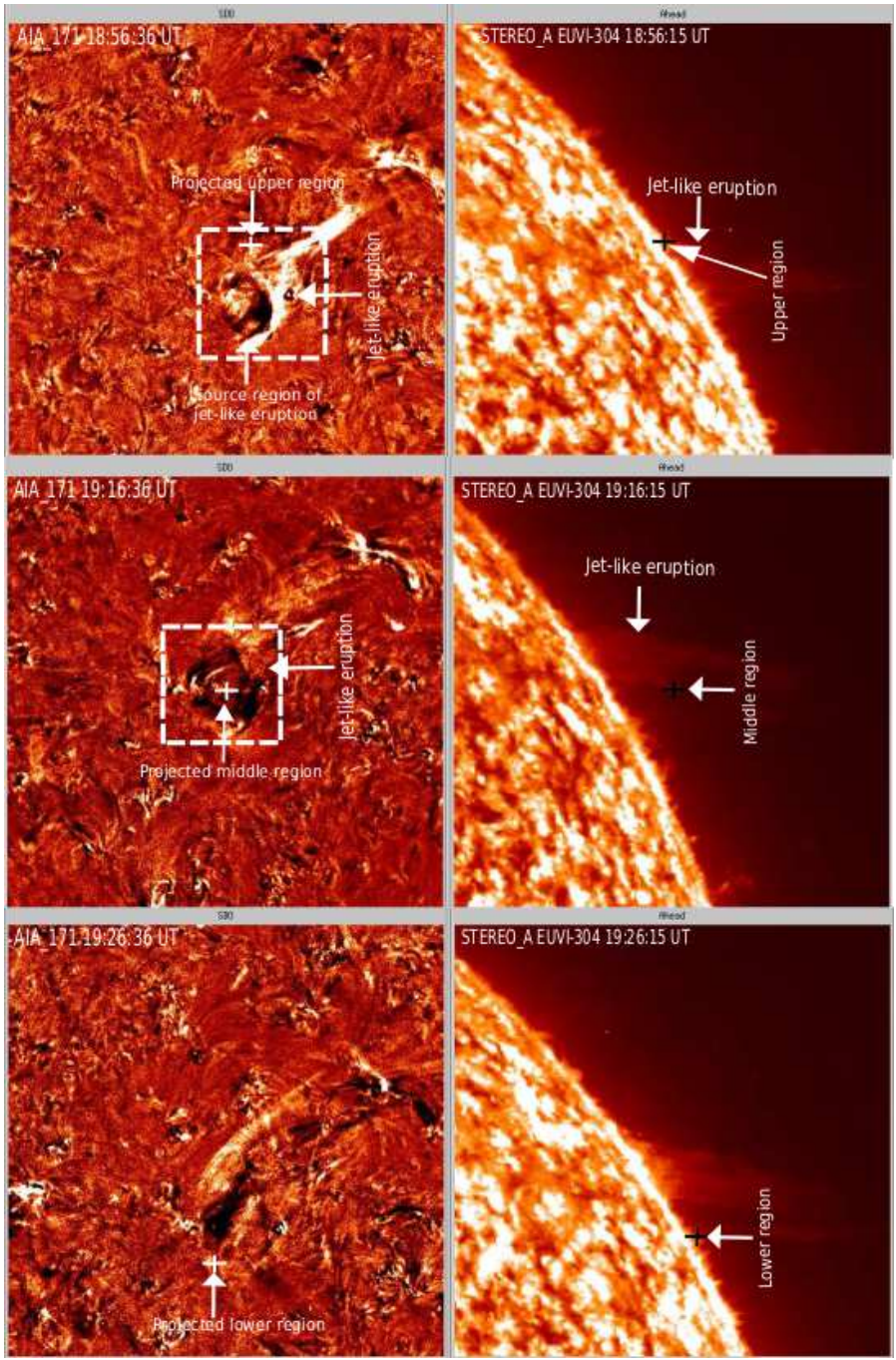}
				\caption{The region of interest (ROI) to observe the source location of the rotating jet-like eruption as observed by SDO/AIA 211 {\AA} and STEREO-A/EUVI-304 {\AA} on 20 August 2018. We adopt the tie-pointing method to locate the source region of the rotating jet-like eruption in the projected plane (EUVI data). We tracked the top, middle, and bottom part of the jet-like eruption in SDO/AIA 211 {\AA} and STEREO-A/EUVI-304 {\AA} (\textit{plus sign in both panels}) by the triangulation technique. The \textit{plus sign} indicates the corresponding location of the jet-like eruption in EUVI-304 {\AA} projected image data. The ROI for the \textit{right panel} corresponds to Figure~4 and ROI for the \textit{left panel} is plotted in Figure~5.}
\end{figure*}
To understand the source region of the flux rope, rotating jet-like eruption, and their associated coronal mass ejections (CMEs), we use the tie-pointing reconstruction technique (Inhester, 2006) to observe the source location of the flux rope on the solar disk first. We used STEREO-A/EUVI 171 and 304 and AIA 171 {\AA} and 193 {\AA} data to observe the source location of the eruption. The position of the two spacecraft (STEREO-A/EUVI and SDO/AIA) was used  as the two viewpoints of the two observers. Any selected point on or above the Sun surface of SDO/AIA data gives the projected epipolar plane on the STEREO-A/EUVI solar imaging data. The epipolar plane has been identified in the two images obtained by two spacecraft STEREO-A/EUVI and SDO/AIA to locate the source region of the eruption. A 3D reconstruction has been possible by tracking the line-of-sight ray that belongs to SDO/AIA images and may be traced-back into the STEREO-A/EUVI images. The intersection of these two traced lines is lying on the same epipolar plane, therefore they give the possible source region of the eruptions. \\
 \begin{figure*}
 \hspace{-1.0cm}
\includegraphics[scale=1.0,angle=0,width=13.0cm,height=13.0cm,keepaspectratio]{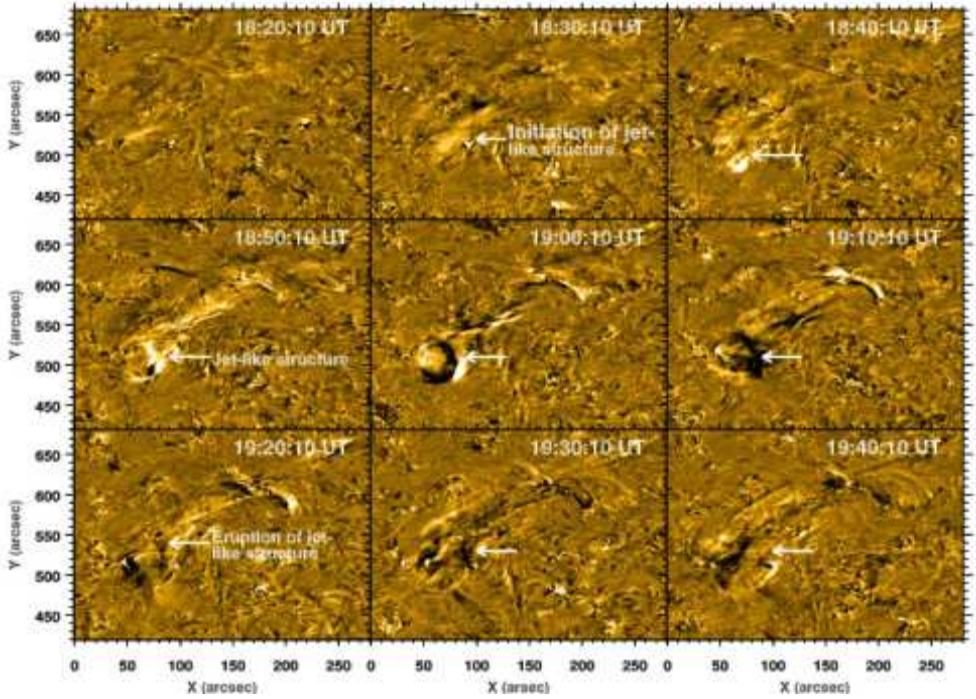}
\caption{The region of interest to observe the jet-like eruption as displayed in the difference images of SDO/AIA 171 {\AA}. The difference images show that a ring-shaped rotating jet-like eruption occurred from the northward coronal hole (CH1).}
\end{figure*}

We used \textit{Wind} spacecraft data (\textit{e.g.} Bougeret {\it et al.}, 1995; Farrell {\it et al.}, 1995; Leinweber, Russel, and Angelopoulos, 1995) to measure the average magnetic field [\textit{\textbf B}] and its components [$B_{x}$, $B_{y}$, $B_{z}$] at one-minute interval (Figure 12). This plot show the effect of CME on Earth's magnetosphere (Figure 12). We obtained the data from the World Data Center for Geomagnetism, Kyoto University, Japan (\urlurl{wdc.kugi.kyoto-u.ac.jp/dst\_realtime/201808/}) with an one-hour interval. We also measure the disturbances in the Earth's magnetic field, which correlate with the Kp-\textsf{indices}. The Kp-\textsf{indices} are measured from the NOAA Space Weather Prediction Center (NOAA/SWPC) with a three-hour temporal interval. Details of the associated results will be explained in the forthcoming sub-sections of article.
 \begin{figure*}
 \hspace{-1.0cm}
\includegraphics[scale=1.0,angle=0,width=13.0cm,height=13.0cm,keepaspectratio]{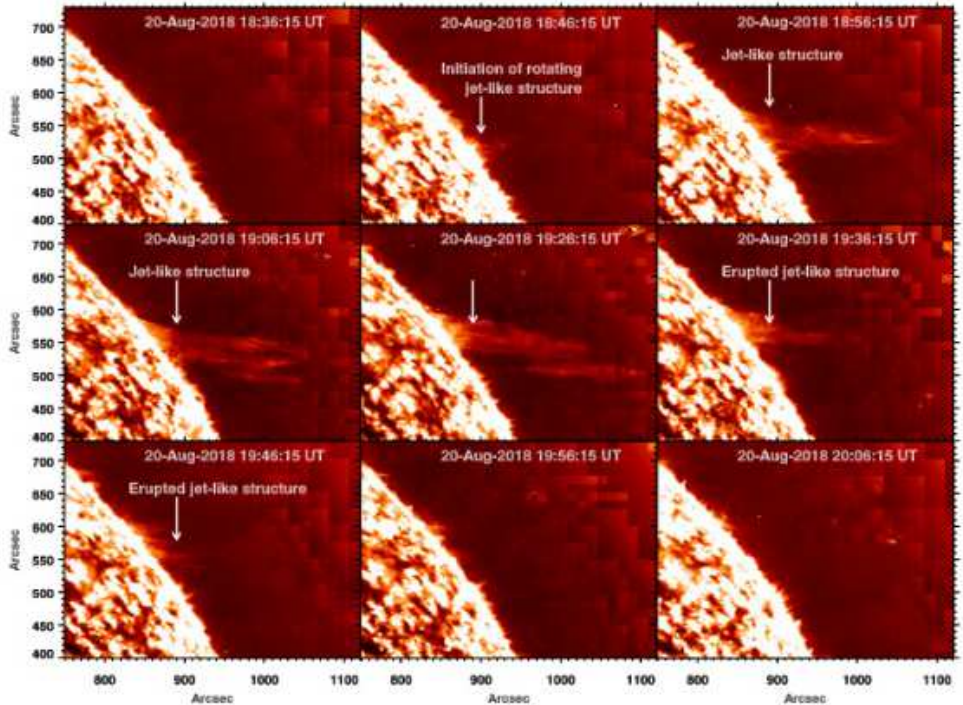}
\caption{The same region of interest as Figure~4 in the projected plane observed by STEREO-A/EUVI-304 {\AA}. Initiation grew and an eruption of the jet-like structure was observed in the STEREO-A/EUVI-304 {\AA} data. }
\end{figure*}

\section{Observational Results}
\subsection{Observation from the {\it Global Oscillation Network Group} (GONG)/BBSO} 
We used GONG/BBSO H$\alpha$ data to observe the partial eruption of the filament. A sequence of temporal images has been used to observe the partial eruption of this quiescent filament (Figure~1). The observed filament moves toward the solar north-west direction on 19 August 2018 and passes over the coronal hole (CH1) on 20 August 2018. As the filament passes over the coronal hole, the open field lines of the coronal hole may be reconnected with it. This reconnection of the open field lines from the coronal hole and overlying filament may be responsible for its partial eruption. Similar dynamics have been discussed by Bhatnagar (1996) and he suggested that a filament eruption occurred from the coronal hole and may be responsible for a strong interplanetary scintillation.
 \begin{figure*}
\includegraphics[scale=1.0,angle=0,width=18.0cm,height=18.0cm,keepaspectratio]{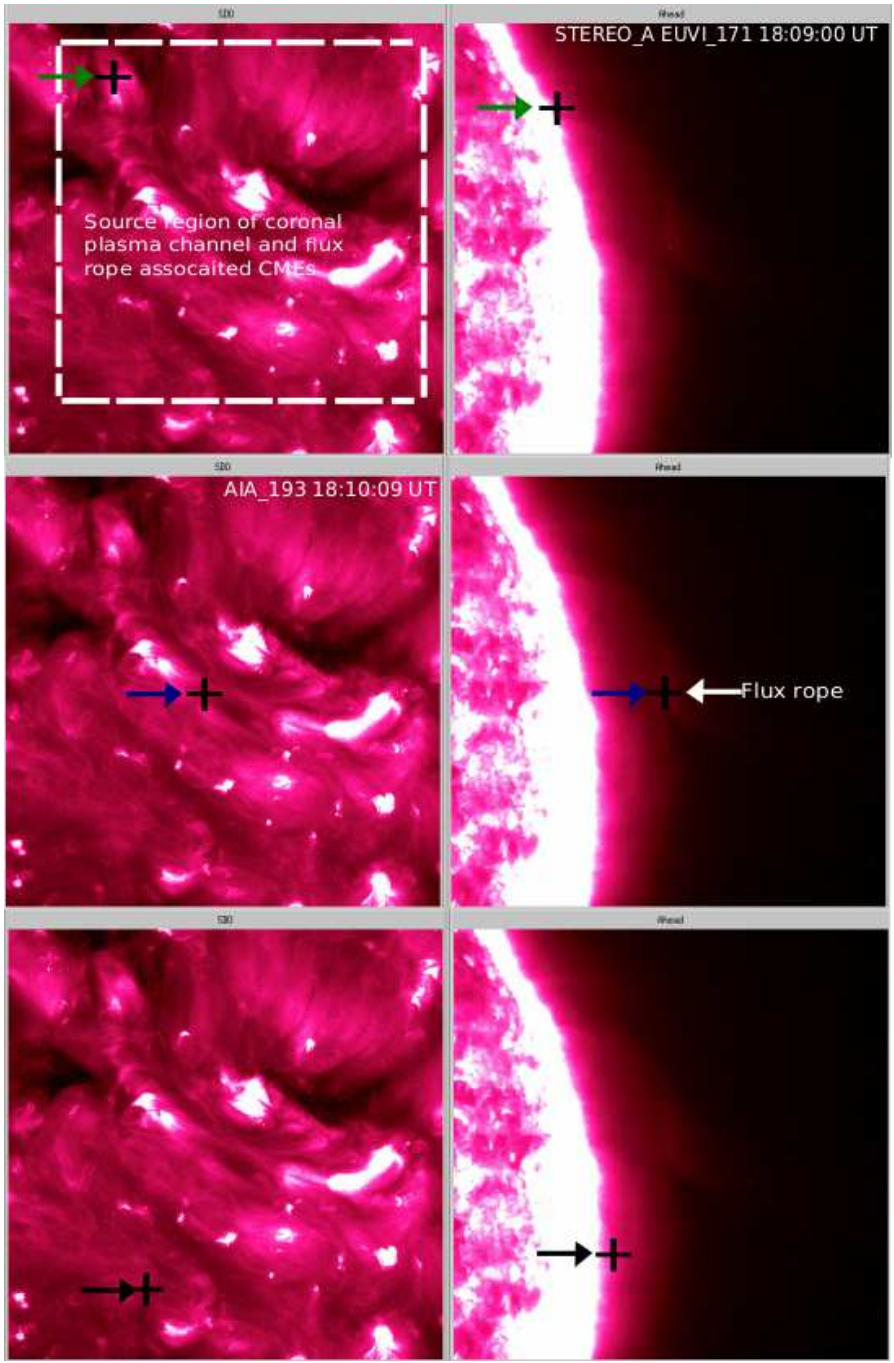}
				 \caption{The region of interest (ROI) to observe the source location of the flux-rope eruption using SDO/AIA 211 {\AA} and STEREO-A/EUVI-171 {\AA} imaging data on 20 August 2018. The tie-pointing method has been used to approximate the source region of the flux-rope eruption on the solar disk (SDO/AIA 211 {\AA}). We tracked the top, the middle, and the bottom leg of the flux rope in the  STEREO-A/EUVI 171{\AA} and SDO/AIA 211{\AA} (\textit{green arrow} for the upper leg, the \textit{blue arrow} for a middle part, and \textit{black arrow} for the bottom leg of the flux rope) by triangulation technique. A diffuse and thin flux rope lying high in the lower corona (STEREO-A/EUVI) is situated above the CPC as observed on the solar disk (SDO/AIA data). The spatial extent of left panel is same as Figure~7.}
\end{figure*}
\subsection{The Eruption of Flux Rope and Jet-Like Structure}
We observe the development of the coronal plasma channel (CPC) on 20 August 2018 using 16 hours of observational data of the SDO/AIA. The coronal plasma channel (CPC) is defined as the confined hot coronal plasma, trapped in the field lines connecting two trans-equatorial coronal holes (CH1 and CH2) and their outer peripheries. It is analogous to the filament channel and may be an extended part of it. However, it does not contain any cool plasma visible in the TR and H${\alpha}$, instead it consists of comparatively hot coronal plasma and related emissions (Figure~2; white-box region). Our observed CPC is similar to the magnetic structure reported by Pevtsov {\it et al.} (2012), which they described as a ``filament channel without filament''. But we abbreviate it as ``coronal plasma channel (CPC)'', because its plasma is visible only at coronal temperatures (Figure~2; white-box region). The hot CPC is visible in all seven EUV filters of SDO/AIA. The coronal plasma channel does not contain cool plasma but it may be observed in terms of magnetic field lines. The disappearance of the cool plasma in the CPC may be due to the slow rise of the magnetic structure of the filament channel. The expansion of the magnetic structure of the coronal plasma channel decreases density and no continuous mass supply occurred in the channel. It connects two trans-equatorial coronal holes situated in the northern and southern hemispheres of the Sun near the central meridian. Initially, the coronal plasma channel (CPC) is in the equilibrium for more than two days. The eruption of CPC is not observable in the SDO/AIA FOV, which is most likely due to the lifting of thin flux rope or it may be situated high in the lower corona. However, the effect of this CPC eruption can be observed (Figure~2). The spreading of this coronal plasma channel (bright strips on CPC system) and post-eruption arcade indicate that an eruption may have occurred from this region forming the aforementioned diffuse CME. This CME has been reported in the CACTUS (\urlurl{secchi.nrl.navy.mil/cactus/}) catalog at 08:09 UT 20 August 2018 in the STEREO-A/COR2 FOV. \\
The spreading of the CPC evolves temporally and spatially. Later, these spreading CPC features (bright strip) interact with the northward lying coronal hole (CH1). It may trigger a rotating jet-like structure from the coronal hole (CH1) (Figures~2\,--\,5). We observe that a rotating jet-like eruption occurred when cool filament passed over the coronal holes. Another possibility to trigger the rotating jet-like eruption is the interaction between open field lines of the coronal hole (CH1) and overlying filament (Figures~2\,--\,5). The overlying filament may reconnect with the open field lines of the coronal hole and may trigger the eruption. However, the main cause of this eruption is not clear. We conjecture that spreading CPC reconnects with the CH1 field lines and form a rotating jet-like structure above the coronal hole (Figures~2\,--\,5). The rotating jet-like structure is observed in seven EUV filters of SDO/AIA and projected STEREO-A/EUVI-304. It is associated with a ring-like rotating CME with a low coronal signature in the inner corona (Figure~8). Using SDO/AIA and STEREO-A/EUVI-304 images, we approximate the location of the jet-like eruption (Figure~3). Above the CPC, a flux rope has also evolved and it erupted at 20:00 UT on 20 August 2018 after the eruption of the CPC (Figures~5\,--\,7). The eruptive CPC is associated with a diffuse CME, which is initially observed on August 20 2018 at 08:09 UT in STEREO-A/COR2 FOV (Figure~8).\\
 \begin{figure*}
\hspace{-1.0cm}
\includegraphics[scale=1.0,angle=0,width=13.0cm,height=13.0cm,keepaspectratio]{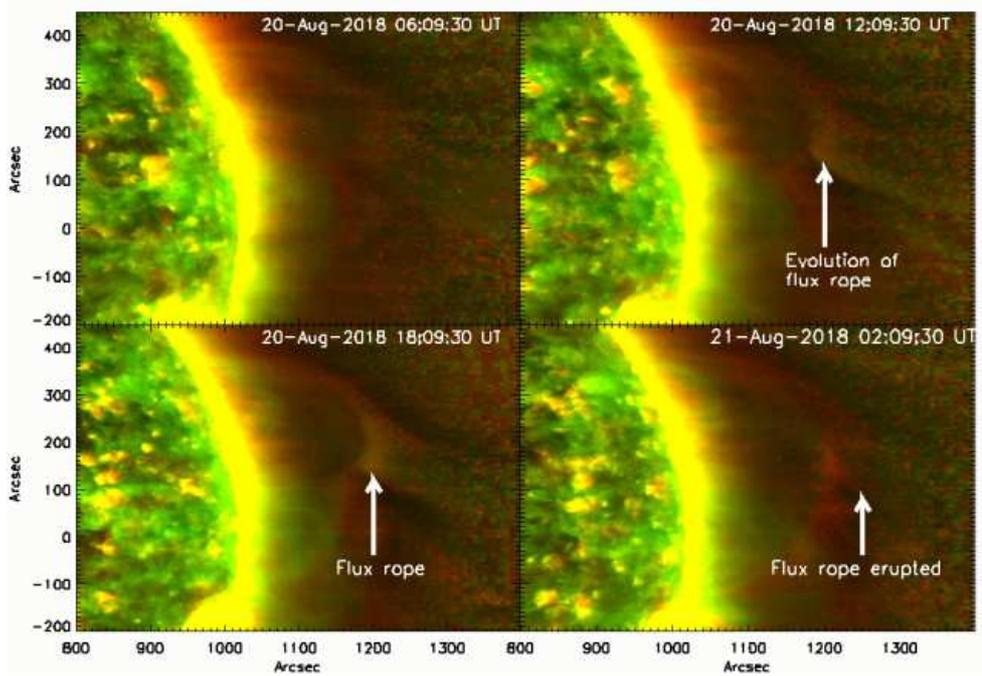}
				 \caption{The region of interest is displayed in the STEREO-A/EUVI (171+195) {\AA} composite images. The composite images observed the dynamics of thin and diffuse flux rope lying high in the lower corona. The top portion of the flux rope has erupted out and the lower part is connected with the solar disk (\textit{lower panel}).}
\end{figure*}
 The evolution of the flux rope and its eruption is not observed in the SDO/AIA FOV. The flux rope is very thin, diffused, and situated high in the lower corona (Figures~6\,--\,7). After some careful investigation, we observed the development of a diffuse and thin flux rope in the STEREO-A/EUVI hotter filters (EUVI-171 and EUVI-195). It consists of a flux-rope associated CME, which is observed into the STEREO-A/COR2 FOV. Initially, this CME starts slowly with a low coronal signature (not observed in the SDO/AIA FOV and having a low signature in STEREO-A/EUVI FOV). With the addition of the EUVI imager and SDO/AIA data, we locate the source region of this eruptive flux rope. We used the tie-pointing method to approximate the location of the flux rope (Figures~6\,--\,7). However, the exact location of the foot-point of the flux rope is not clear on the solar disk in the SDO/AIA data. We used a composite image of STEREO-A/EUVI-171 and EUVI-195 to observe the initiation, growth, and eruption of this diffused flux rope (Figure~7). Initially, we observe that there is no signature of overlying flux situated high in the corona. Legs of the flux rope are observed into the hotter filters of STEREO-A/EUVI-171 and EUVI-195. We found a very faint and diffuse flux rope that has been evident over the western limb. Later, the top portion of the flux rope erupted and the lower part of the flux rope is connected with the solar limb (Figure~7, lower panel). A flux-rope-associated CME has been evident in the outer corona as observed in the STEREO-A/COR2 FOV with a low coronal signature in the lower corona. 
\begin{figure*}
\hspace{-1.0cm}
\includegraphics[scale=1.0,angle=0,width=13.0cm,height=13.0cm,keepaspectratio]{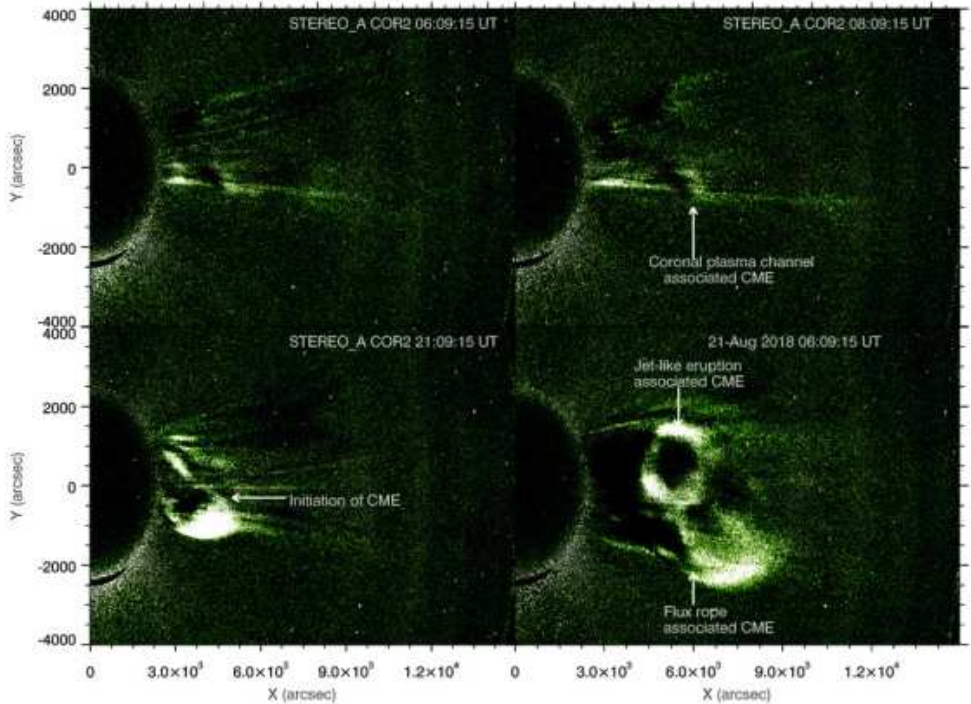}
\caption{Half FOV of STEREO-A/COR2 to observe the eruption of the CMEs. A coronal-plasma channel associated CME appeared first with its low coronal signature (upper panel). The CMEs associated with the flux rope and rotating jet-like eruption have been observed using STEREO-A/COR2. Later these CMEs merge within each other and propagate towards the outer corona in the form of a compound CME.}
\end{figure*}
\section{Behaviour of CMEs from Outer Corona to the Interplanetary Space}
We observe three CMEs on 20 August 2018 using STEREO-A/COR2, HI1 and HI2 data. Initially, a coronal-plasma-channel associated eruption is responsible for the stealth CME, which is firstly evident at 08:09 UT on 20 August 2018 in STEREO-A/COR2 FOV. The eruption of CPC has not been observed into the lower corona; however, its consequences may be observed in form of spreading coronal plasma channel and post eruption-arcade. Later, a faint and slow-rising flux rope-like structure developed in EUVI-A FOV above the west limb (Figures~6\,--\,7). This flux rope has developed above the coronal plasma channel (Figure~6) and erupted around 20:09 UT. The eruptive flux rope does not show any sign in the STEREO-A/COR1 FOV and very weak signature in the lower corona. This fact indicates the stealth nature of the CME. However, the post-eruption arcade and legs of the flux rope are observed in the EUVI-A FOV (Figure~7). Observations of EUVI-A/171{\AA} and 195{\AA} suggest that the apparent breakup of the top portion of the flux rope between 18:09 UT to 22:09 UT (Figure~7) has occurred. The eruption of the coronal plasma channel may also be responsible for the eruption of a jet-like structure from the coronal hole (Figures~2\,--\,4). The eruptive flux rope (Figure~7) and jet-like structure (Figures~3\,--\,5) are associated with two CMEs as observed in the STEREO-A/COR2 FOV. These two CMEs first appeared in the COR2A FOV at 20:09 UT (Figure~8). They grew spatially into the outer corona and merge within each other. These CMEs over expanded and flattened as they passed through the outer corona. The over expansion of the CMEs is a rare phenomenon in the outer corona (Vourlidas {\it et al.}, 2011). The compound CME has larger angular width, thus it may be Earth-directed. Depending upon the surrounding magnetic field, the CME may be expanded and accelerated as they pass through the outer corona (Cargill {\it et al.}, 1999). We observed that the flux rope-associated CME and the rotating jet-like eruption associated CME merge within each other (Figure~8, COR2 FOV). The merged CME consists of larger angular width and bigger size (Figures~8\,--\,9). We abbreviate it as an overexpansion of the compound CME qualitatitvely as detected in the given observational base-line.  \\
\begin{figure*}
\hspace{-1.0cm}
\includegraphics[scale=1.0,angle=0,width=13.0cm,height=13.0cm,keepaspectratio]{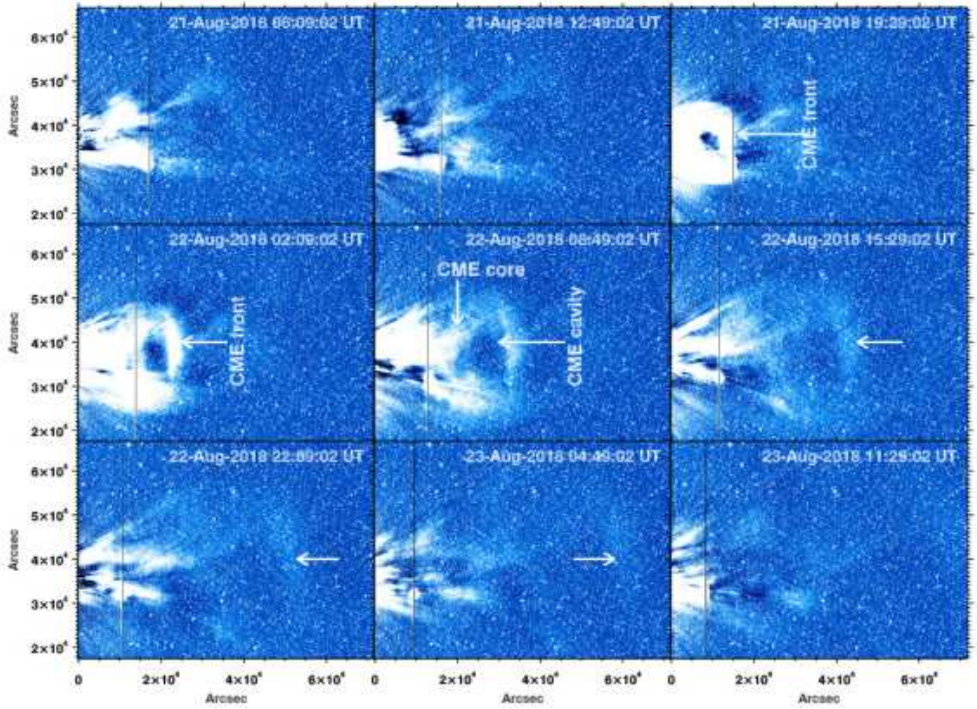}
\caption{Formation of the merged CME in interplanetary space  as observed by STEREO-A/HI1. The compound CME has larger angular width in interplanetary space and propagated towards the Earth. It crossed the lower interplanetary space on 23 August 2018.}
\end{figure*}
This expanded, compound CME further passes through the low interplanetary space as observed by STEREO-A/HI1 (Figure~9). The CME crosses the HI1 FOV (15\,--\,90\,R$_\odot$). The CME first appears at 05:09 UT on 21 August 2018. We use an array of images to observe the full development of the CME into low interplanetary space. Before the expansion of the CME, a weak CME front (without its core) moves ahead, which is generated due to the ejection of the upper part of the flux rope into the Sun's corona. The CME front, followed by the cavity region and core of the CME has been indicated by white arrows (Figure.~9). It crosses the low interplanetary space after 10:30 UT on 23 August 2018 (lower panel, Figure~9). It further passes through the interplanetary space as observed by STEREO-A/HI2 (90\,R$_\odot$\,--\,4 AU). 
\\

Figure~11 displays the aligned images of STEREO-A/COR2 (2.5 to 15 solar radii), Heliospheric Imager (HI)-1A (15 to 80 solar radii), and Heliospheric Imager (HI)-2A (up to 4\,AU, 1\,AU=1.49$\times$10$^{11}$ m Sun--Earth distance) on 26 August 2018 at 16:09 UT covering from the Sun’s outer corona to the Earth and beyond. We have taken the array of images of STEREO-A/HI2 from RAL SPACE STEREO (\urlurl{www.stereo.rl.ac.uk/}) to observe the CME in the interplanetary space at 1\,AU (Figure~10). This figure clearly shows that the confined CME overexpanded into the outer corona (COR2 FOV) and it further expands and moves in the heliosphere towards 1\,AU (HI-1A and HI-2A FOV) over a trajectory. The CME has initially observed on 24 August 2018 at 08:09 UT as observed into the STEREO-A/HI2 FOV. We observe that a diffuse CME passes through interplanetary space, which propagates towards 1\,AU. The upper part of the CME is not interacting with Earth magnetosphere as it is deflected in other directions (Figure~11). The lower part of the CME interacts with the Earth's magnetosphere on 26 August 2018 at 1\,AU. The interaction of this diffuse CME is responsible for an intense geomagnetic storm. 
\begin{figure*}
\hspace{-1.0cm}
\includegraphics[scale=1.0,angle=0,width=13.0cm,height=13.0cm,keepaspectratio]{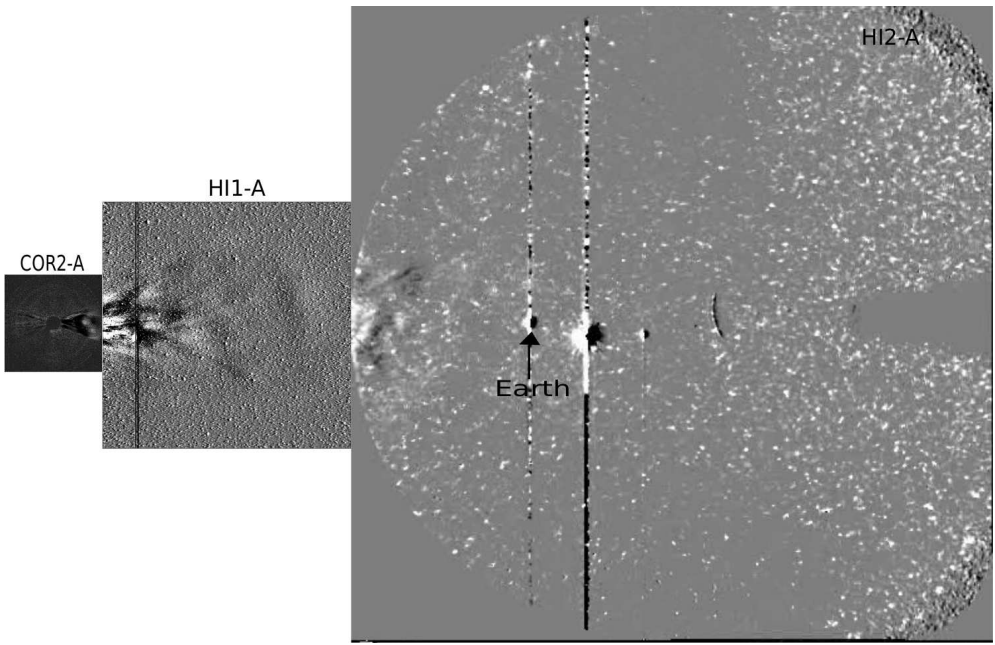}
				\caption{Aligned and composite images of STEREO-A/COR2 (2.5\,--\,15\,R$_\odot$, \textit{Heliospheric Imager} (HI1; 15\,--\,90{\(R_\odot\)}), and \textit{Heliospheric Imager} (HI2; up to 4\,AU) onboard the \textit{Solar TErrestrial Relations Observatory} (STEREO) on 26 August 2018 at 10:09 UT, display the interaction of CME with the Earth's magnetosphere.}
\end{figure*}

\section{Intense Geomagnetic Storm Due to the Observed CMEs}
We observe the propagation of the compound/stealth CME at 1\,AU and its impact on the Earth's magnetosphere. The interaction of the CME has been observed near the Earth's by using STEREO-A/HI2 and simultaneously measureing the Dst index. The Dst index started to decrease after 12:00 UT on 25 August 2018 as the middle part of the CME reached near to the Earth's magnetosphere. The main phase of the geomagnetic storm is indicated by a dotted line. The Dst index reaches negative values up to -175 nT (Figure~12) and exhibits an intense geomagnetic storm at 10:09 UT on 26 August 2018. The recovery phase is indicated by arrows (Figure~12; top panel) and it did not recover up to 29 August 2018. Figure~12 (middle and lower panel) displays the interplanetary magnetic field [\textit{\textbf B}] and the component of the magnetic field [$B_{x}$, $B_{y}$, and $B_{z}$]. The $B_{x}$- and $B_{y}$-components of the interplanetary magnetic field are not important for geomagnetic activities. The $z$-component of the interplanetary magnetic field [$B_{z}$] tends to negative values up to $\approx$-18 nT after the arrival of this diffuse CME near Earth's magnetosphere. The negative increment of the Dst index and $z$-componet of the interplanetary magnetic field during this interaction, indicate the onset of the geomagnetic storm. The increment in the plasma density [$N_\mathrm{p}$] indicates the shock signature of the coronal mass ejection. We observe that the plasma density remains constant. The magnitue of the velocity near the Earth magnetosphere is $\approx$350 km\,s$^{-1}$ (Figure~12; lower panel), which indicates that the stealth CME is moving with slow velocity in the interplanetary space. We also examine the Kp index, which quantifies the disturbance in the Earth's magnetic field. The Kp index lies between integers 0\,--\,9 with less than 5 indicating calm weather conditions and 5 or more indicating a geomagnetic storm. We obtain the Kp indices from the NOAA Space Weather Prediction Center (NOAA/SWPC) on 25 August 2018 to 28 August 2018. We observe that Kp index is less than 4 at 00:00 UT on 26 August 2018. Later, we conjecture that the Kp index became more than 5 and it reaches up to 7 on 26 August 2018 at 06:00\,--\,09:00 UT. It remains more than 5 on the same days and later it decreases (Figure~13). We use the STEREO-A/HI2 imaging data, measured Dst index, and simultaneously estimated Kp index from 25\,--\,28 August 2018. These simultaneous observations indicate that an intense geomagnetic storm has occurred on the same day. The recovery phase starts from 27 August 2018 which may be observed from estimating the Dst and Kp indices (Figures~12\,--\,13).
\begin{figure*}
\hspace{-0.5cm}
\includegraphics[scale=1.0,angle=0,width=13.0cm,height=13.0cm,keepaspectratio]{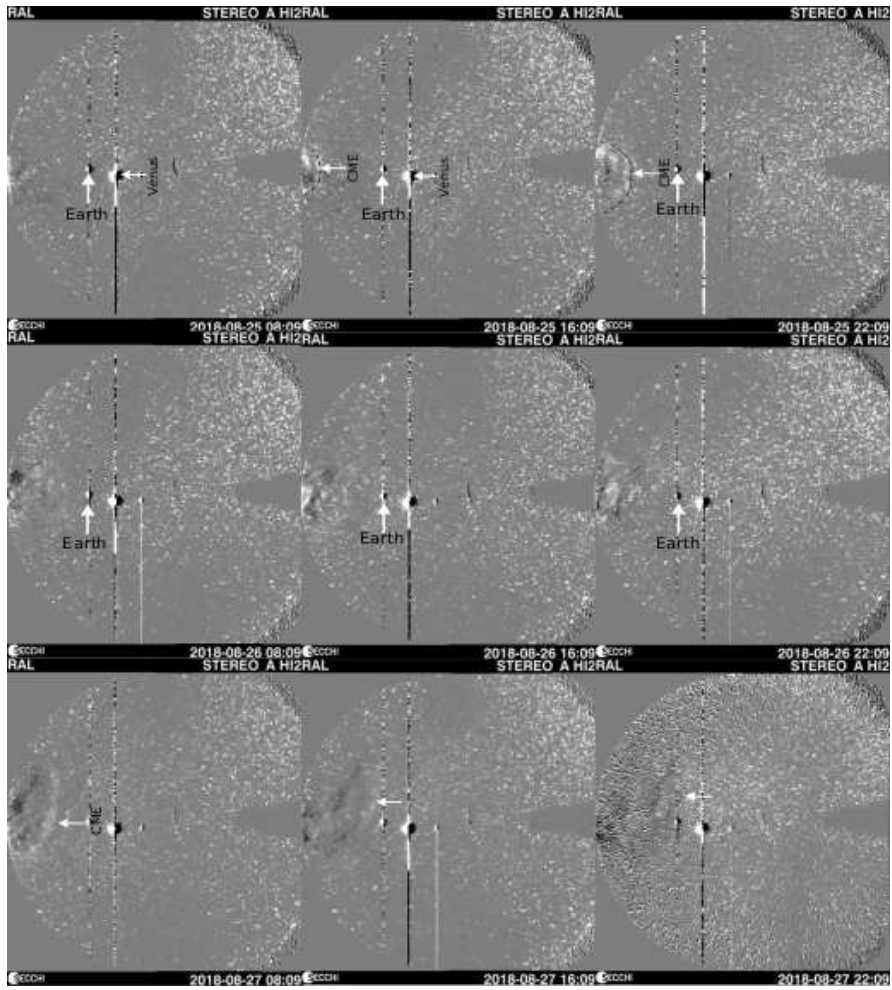}
\caption{The propagation of collective CMEs towards 1\,AU and interaction with the Earth’s magnetosphere. The sequence of STEREO-A/HI2 images was used to observe the merged CME in interplanetary space. The observed CME moves toward 1\,AU and interacts with the Earth's magnetosphere on 26 August 2018 and creates large-scale disturbances.}
\end{figure*}
\section{Discussion and Conclusions}
\begin{figure*}
 \hspace{-1.0cm}
\includegraphics[scale=1.0,angle=0,width=14.0cm,height=14.0cm,keepaspectratio]{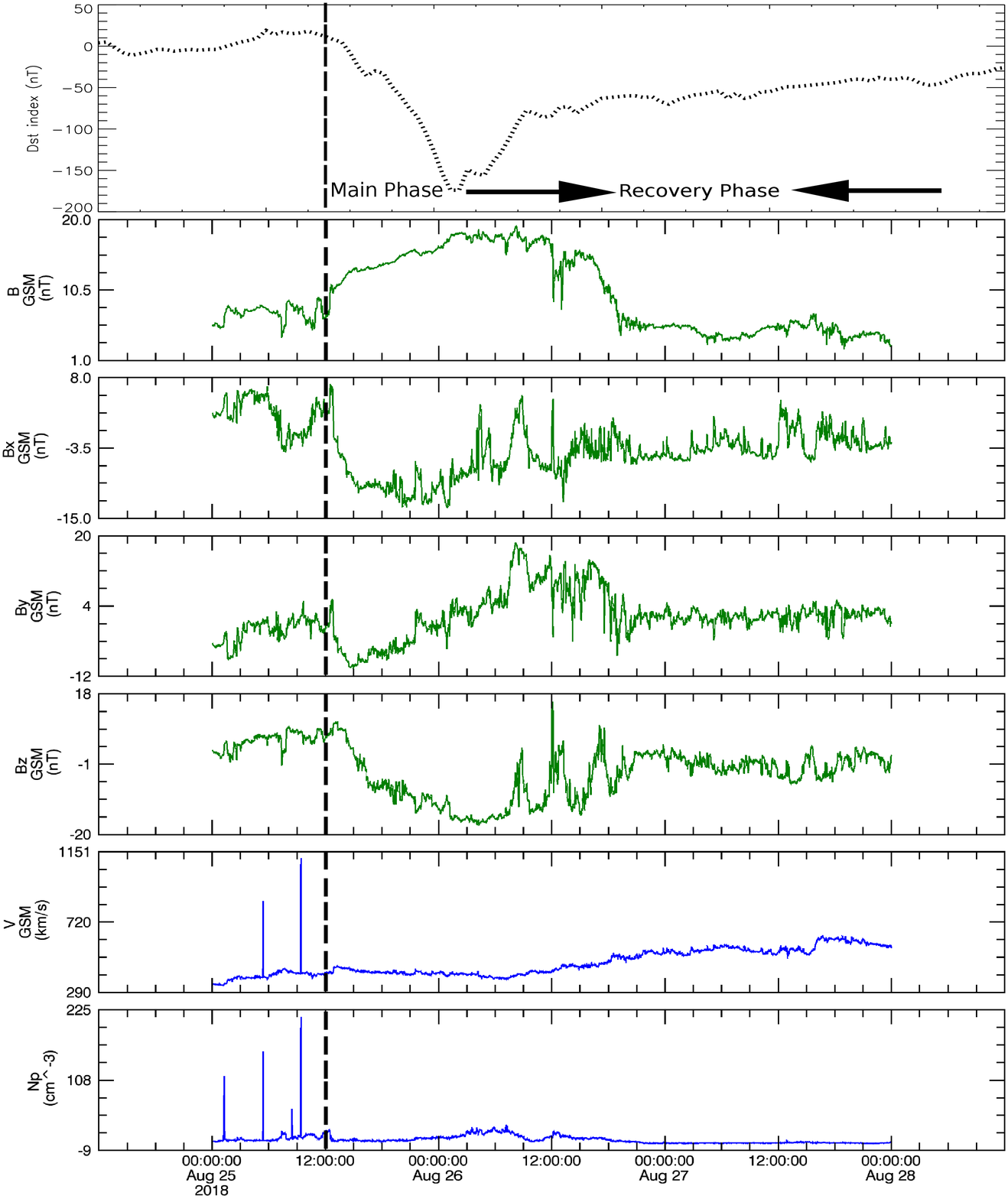}
				\caption{Dst index after the simultaneous arrival of CMEs. The Dst index has started to decrease from positive to negative values. The Dst index decreases to -176\,nT at 12:09 UT on 26 August 2018 exhibiting an intense and long duration geomagnetic storm. The average magnetic field [\textit{\bf B}] and the component of the magnetic field [$B_{x}$, $B_{y}$, and $B_{z}$ ], plasma flow speed [{\bf V}], and proton density [$N_\mathrm{p}$]. The \textit{black-dotted vertical lines} show the arrival time of the CME at 1\,AU at 12:09 UT on 25 August 2018. The $z$-component of the magnetic field, Dst index, and other parameters simultaneously show the disturbances in the Earth magnetosphere. }
\end{figure*}
A partially erupting quiescent filament is associates with a coronal plasma channel (CPC) on 20 August 2018. The CPC connects two trans equatorial coronal holes (CH1 and CH2) and erupted on 20 August 2018 with its low coronal signature. The spreading CPC reconnects with the open field lines of a coronal hole (CH1) and triggered a rotating jet-like eruption. Later, a thin and diffuse flux rope has developed above the CPC and erupted with very faint coronal signature. These eruptions are associated with the stealth type CME. The compound stealth CME interacts with Earth's magnetosphere on 26 August 2018 and produce a third intense geomagnetic storm of Solar cycle 24. \\
A quiescent filament partially erupted on 20 August 2018 at 18:30 UT when it passes over the northward coronal hole (CH1). A coronal plasma channel is developed between the two coronal holes (CH1 and CH2; Figure.~2). It grew spatially and temporally for more than two days and erupted after 05:00 UT on 20 August 2018. It is difficult to observe how and when the coronal plasma channel eruption started to occur, but the spreading coronal plasma-channel and post-eruption arcade indicate that an eruption has already begun from the same region (Figures~2\,--\,4). This eruption was later seen in STEREO-A/COR2 images as an Earth-directed CME at 12:00 UT (Figure~8), however, its traces were very faint in the lower corona. The spreading of coronal plasma-channel and filament lying over the coronal hole may interact with its open-field lines, and trigger the rotating jet-like structure. The rotating jet-like eruption is observed in all seven AIA EUV filters and STEREO-A/EUVI-304 {\AA}. Above the coronal plasma channel, a very thin and diffuse flux rope also developed, which is not observed in the SDO/AIA FOV, but after some inspection we find a thin and diffuse flux rope lying high in the lower corona (Figures~6\,--\,7). The top portion of the flux rope has erupted and the lower part of the flux rope is connected with the solar disk. The location of the overlying diffuse flux rope is approximated by using tie-pointing methods. We find that the location of the flux rope is lying above the coronal-plasma channel, which connects the two coronal holes (CH1 and CH2; Figure~2). Later, we observed that three consecutive CMEs were evident in the outer corona with very faint evidence in the lower corona. The CPC eruption is associated with the first stealth CME as evident at 08:30 UT (Figure.~8; upper panel). It is followed by a slow rising rotating jet-like CME, and thereafter a flux rope associated CME (Figure.~8; lower panel). 
In the outer corona, we observe  well-developed signatures of these CMEs. These three CMEs further merge within each other and form a bigger compound CME, which passes through the interplanetary space (Figures~9\,--\,11).\\

\begin{figure*}
\hspace{-1.0cm}
\includegraphics[scale=1.0,angle=0,width=13.0cm,height=13.0cm,keepaspectratio]{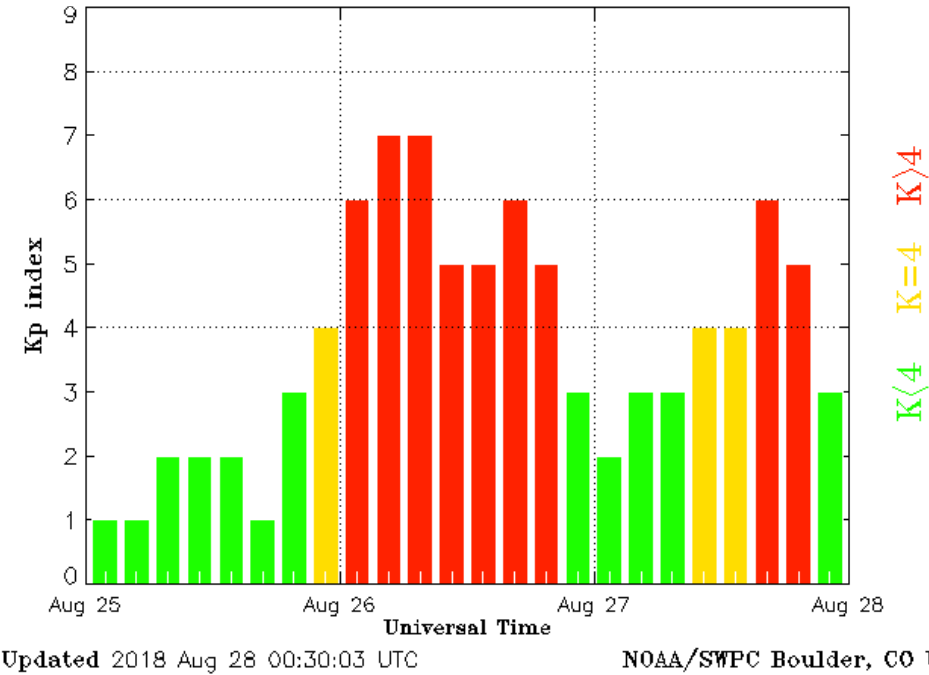}
\caption{Kp index (disturbances in the Earth's magnetic field) from 25 August to 28 August 2018. The Kp index reaches up to 7, which indicates an intense geomagnetic storm at 12:00 UT on 26 August 2018. }
\end{figure*}
The compound CME formed by the merging of the three stealth CMEs crosses the low interplanetary space on 23 August 2018 as observed by STEREO-A/HI1 (Figure~9). This bigger CME has been observed near the Earth at 1\,AU after its arrival there. We observe that CME associated interplanetary magnetic field interacts with the Earth's magnetosphere on 25 August 2018. Interaction of CME with the Earth’s magnetosphere produces ring-current and causes the geomagnetic storm (\textit{i.e.} depression in Dst index) on 26 August 2018 (Figures~12\,--\,13). We observe the interaction of CME near the Earth magnetosphere and simultaneously measure the Dst index, $z$-component of interplanetary magnetic field, and Kp index. The imaging data and three parameters (\textit{\textbf B}, $B_{z}$, and Kp-index) indicate that an intense geomagnetic storm appeared on 26 August 2018 (Figures~12\,--\,13). Interplanetary magnetic field (IMF) dragged by the solar wind and CME system may become strictly southward at a particular instance, and thus may create an intense geomagnetic storm at the Earth (Dst= -176 nT). Earth’s magnetosphere may try to alter the interplanetary magnetic field [$B_{z}$] both in the magnitude and direction by their field, thus creating a strong geomagnetic storm. Simultaneous measurement of the K index (disturbances in the horizontal component of the Earth's magnetic field) indicates that it reaches more than 5, which suggests the third strongest geomagnetic storm of the Solar Cycle 24 (Figure~13). \\

The solar filament may reconnect with the open-field line of the coronal hole and responsible for the filament eruption (Bhatnagar, 1996). Similar physical processes have been observed for the rotating jet-like eruption from the coronal hole. These eruptions (coronal plasma channel, rotating jet-like, and flux rope-associated eruption) have a very low/faint signature in the inner corona (\textit{e.g.} Robbrecht, Patsourakos, and Vourlidas, 2009; Nitta and Mulligan, 2017; Gopalswamy {\it et al.}, 2009; Nieves-Chinchilla {\it et al.}, 2013; D'Huys {\it et al.}, 2015; Lynch {\it et al.}, 2016). The stealth-type CMEs may erupt from the filament channel even without filament eruption and could be responsible for intense geomagnetic activity (Pevtsov {\it et al.}, 2012). Bhatnagar (1996) suggests that filament eruption occurs from coronal holes, further giving rise to a strong interplanetary impact on the Earth's magnetosphere. We observe that the Dst index reaches a negative value of -176 nT (Figure~12) and exhibits a strong geomagnetic storm and simultaneously the $z$-component of the interplanetary magnetic field decreases to -18 nT negative value (Yokoyama and Kamide, 1997; Gonzalez and Echer, 2005). Richardson and Cane (2011) discussed  statistically the relationship between Dst index and Kp during the Solar Cycle 23. The negative value of the Dst index is directly proportional to the Kp indices. Similar physical processes occurred during the stealth CME that erupted on 20 August 2018 as observed in the present article. The present case describes the formation of the stealth CMEs from the coronal hole and CPC, which collectively triggered the strong geomagnetic storm on 26 August 2018.\\

\section{Acknowledgments}
S. Mishra and A.K. Srivastava
acknowledge the DST-SERB (YSS/2015/000621) project. S. Mishra thanks the Department of Physics, Indian Institute of Technology (BHU)
for providing him teaching assistantship and computational facilities. A.K. Srivastava acknowledges the UKIERI Research Grant for the support of his research. The authors acknowledge the GONG/H$\alpha$, SDO/AIA, and STEREO-A/EUVI, COR2, HI1 and HI2 observational data. They also acknowledge the use of Dst index data from the Kyoto University, \textit{Wind} data, and NOAA/SWPC data for the estimation of Kp indices. The reviewer evaluation and 
remarks on the article are thankfuly acknowledged.
This work utilizes data obtained by the \textit{Global Oscillation Network Group}
(\,GONG\,) program, managed by the National Solar Observatory, which is
operated by AURA, Inc. under a cooperative agreement with the National
Science Foundation. The data were acquired by instruments operated by the
Big Bear Solar Observatory, High Altitude Observatory, Learmonth Solar
Observatory, Udaipur Solar Observatory, Instituto de Astrof\'{\i}sica de
Canarias, and Cerro Tololo Interamerican Observatory.
The STEREO/SECCHI data used here were produced by an international
consortium of the Naval Research Laboratory (\,USA\,), Lockheed Martin
Solar and Astrophysics Lab (\,USA\,), NASA Goddard Space Flight Center
(\,USA\,), Rutherford Appleton Laboratory (\,UK\,), University of
Birmingham (\,UK\,), Max-Planck-Institut for Solar System Research
(\,Germany\,), Centre Spatiale de Li\`ege (\,Belgium\,), Institut d'Optique
Thorique et Appliqu\'ee (\,France\,), and Institut d'Astrophysique Spatiale
(\,France\,). The USA institutions were funded by NASA, the UK institutions
by the Science \& Technology Facility Council (\,which used to be the
Particle Physics and Astronomy Research Council, PPARC\,), the German
institutions by Deutsches Zentrum f\"{u}r Luftund Raumfahrt e.V. (\,DLR\,),
the Belgian institutions by Belgian Science Policy Office, and the French
institutions by Centre National d'Etudes Spatiales (\,CNES\,) and the
Centre National de la Recherche Scientifique (\,CNRS\,). The NRL effort was
also supported by the USAF Space Test Program and the Office of Naval
Research.
Data courtesy of NASA/SDO and the AIA science team.

\section{Declaration of Potential Conflicts of Interest}
The authors declare that they have no conflicts of interest.


\begin{thebibliography}{}
				\bibitem[Adams \emph{et al.}(2014)]{2014ApJ...783...11A}Adams, M., Sterling, A.C., Moore, R.L., Gary, G.A.: 2014, {\it Astrophys. J.} {\bf 783}, 11, \doiurl{10.1088/0004-637X/783/1/11}.
				\bibitem[Aly(1990)]{1990CoPhC..59...13A}Aly, J.J.: 1990, {\it Comp. Phys. Comm.} {\bf 59}, 13, \doiurl{10.1016/0010-4655(90)90152-Q.}.
				\bibitem[Alzate and Morgan(2017)]{2017ApJ...840..103A}Alzate, N., Morgan, H.: 2017, {\it Astrophys. J.} {\bf 840}, 103, \doiurl{10.3847/1538-4357/aa6caa}.
				\bibitem[Amari \emph{et al.}(2000)]{2000ApJ...529L..49A}Amari, T., Luciani, J.F., Mikic, Z., Linker, J.: 2000, {\it Astrophys. J.} {\bf 529}, L49, \doiurl{10.1086/312444}.
				\bibitem[Amari \emph{et al.}(2003)]{2003ApJ...595.1231A}Amari, T., Luciani, J.F., Aly, J.J., Mikic, Z., Linker, J.: 2003, {\it Astrophys. J.} {\bf 595}, 1231, \doiurl{10.1086/345501}.
				\bibitem[Antiochos, DeVore, and Klimchuk(1999)]{1999ApJ...510..485A}Antiochos, S.K., DeVore, C.R., Klimchuk, J.A.: 1999, {\it Astrophys. J.} {\bf 510}, 485, \doiurl{10.1086/306563}.
				\bibitem[Archontis and T{\"o}r{\"o}k(2008)]{2008A&A...492L..35A}Archontis, V., T{\"o}r{\"o}k, T.: 2008, {\it Astron. Astroph.} {\bf 492}, L35, \doiurl{10.1051/0004-6361:200811131}.
				\bibitem[Barnes and Sturrock(1972)]{1972ApJ...174..659B}Barnes, C.W., Sturrock, P.A.: 1972, {\it Astrophys. J.} {\bf 174}, 659, \doiurl{1086/151527}.
				\bibitem[Bhatnagar(1996)]{1996Ap&SS.243..105B}Bhatnagar, A.: 1996, {\it Astrophys. Space Sci.} {\bf 243}, 105, \doiurl{10.1007/BF00644039.}.
\bibitem[Bougeret \emph{et al.}(1995)]{1995SSRv...71..231B}Bougeret, J.-L., Kaiser, M.L., Kellogg, P.J., Manning, R., Goetz, K., Monson, S.J.,
				Monge, N., Friel, L., Meetre, C.A., Perche, C., Sitruk, L., Hoang, S.: 1995, {\it Space Sci. Rev.} {\bf 71}, 231, \doiurl{10.1007/BF00751331}.
\bibitem[Cargill \emph{et al.}(2000)]{2000JGR...105.7509C}Cargill, P.J., Schmidt, J., Spicer, D.S., Zalesak, S.T.: 2000, {\it J.  Geophys. Res.} {\bf 105}, 7509, \doiurl{10.1029/1999JA900479}.
\bibitem[Chen(1989)]{1989ApJ...338..453C}Chen, J.: 1989, {\it Astrophys. J.} {\bf 338}, 453, \doiurl{10.1086/167211}.
\bibitem[Chen(2011)]{2011LRSP....8....1C}Chen, P.F.: 2011, {\it Liv. Rev. Solar Phys.} {\bf 8}, 1, \doiurl{org/10.12942/lrsp-2011-1.}.
\bibitem[Chen and Shibata(2000)]{2000ApJ...545..524C}Chen, P.F., Shibata, K.: 2000, {\it Astrophys. J.} {\bf 545}, 524, \doiurl{10.1086/317803}.
\bibitem[Chertok \emph{et al.}(2002)]{2002ApJ...567.1225C}Chertok, I.M., Obridko, E.I., Mogilevsky, V.N., Shilova, N.S., Hudson, H.S.: 2002, {\it Astrophys. J.} {\bf 567}, 1225, \doiurl{10.1086/338584}.
\bibitem[Deng \emph{et al.}(2001)]{2001SoPh..204...11D}Deng, Y., Wang, J., Yan, Y., Zhang, J.: 2001, {\it Solar Phys.} {\bf 204}, 11, \doiurl{10.1023/A:1014258426134}.
\bibitem[Denker \emph{et al.}(1999)]{1999SoPh..184...87D}Denker, C., Johannesson, A., Marquette, W., Goode, P.R., Wang, H., Zirin, H.: 1999, {\it Solar Phys.} {\bf 184}, 87, \doiurl{10.1023/A:1005047906097}.
\bibitem[D'Huys \emph{et al.}(2014)]{2014ApJ...795...49D}D'Huys, E., Seaton, D.B., Poedts, S., Berghmans, D.: 2014, {\it Astrophys. J.} {\bf 795}, 49,\doiurl{10.1088/0004-637X/795/1/49}.
\bibitem[Farrell et al.(1995)]{1995ITM....31..966F} Farrell, W.~M., Thompson, R.~F., Lepping, R.~P., Byrnes, J.~B.\ 1995, IEEE Transactions on Magnetics, {\bf 31}, 966, \doiurl{10.1109/20.364770}.
\bibitem[Filippov and Koutchmy(2008)]{2008AnGeo..26.3025F} Filippov, B., Koutchmy, S.\ 2008, {\it Ann. Geophys.}, {\bf 26}, 3025, \doiurl{10.5194/angeo-26-3025-2008}. 
\bibitem[Gonzalez and Echer(2005)]{2005GeoRL..3218103G}Gonzalez, W.D., Echer, E.: 2005, {\it Geophys. Res. Lett.} {\bf 32}, L18103 , \doiurl{10.1029/2005GL023486}.
\bibitem[Gopalswamy(2004)]{2004ASSL..317..201G} Gopalswamy N. (2004) A Global Picture of CMEs in the Inner Heliosphere. In: Poletto G., Suess S.T. (eds.) \textit{The Sun and the Heliosphere as an Integrated System. Astrophys. Space Sci. Lib.} \textbf{317}, 201. Springer, Dordrecht. \doiurl{10.1007/978-1-4020-2831-9_8}.
\bibitem[Gopalswamy \emph{et al.}(2003)]{2003ApJ...586..562G}Gopalswamy, N., Shimojo, M., Lu, W., Yashiro, S., Shibasaki, K., Howard, R.A.: 2003, {\it Astrophys. J.} {\bf 586}, 562. 
\bibitem[Gopalswamy \emph{et al.}(2009)]{2009JGRA..114.0A22G}Gopalswamy, N., M{\"a}kel{\"a}, P., Xie, H., Akiyama, S., Yashiro, S.: 2009, {\it J. Geophys. Res. A} {\bf 114}, A00A22, \doiurl{10.1029/2008JA013686}.
\bibitem[Gosling(1993)]{1993JGR....9818937G}Gosling, J.T.: 1993, {\it J. Geophys. Res.} {\bf 98}, 18937. 
\bibitem[Green \emph{et al.}(2018)]{2018SSRv..214...46G}Green, L.M., T{\"o}r{\"o}k, T., Vr{\v s}nak, B., Manchester, W., Veronig, A.: 2018, {\it Space Sci. Rev.} {\bf 214}, 46, \doiurl{10.1007/s11214-017-0462-5}.
\bibitem[Harrison(1995)]{1995A&A...304..585H}Harrison, R.A.: 1995, {\it Astron. Astrophys.} {\bf 304}, 585.  
\bibitem[Harvey \emph{et al.}(2011)]{2011SPD....42.1745H}Harvey, J.W., Bolding, J., Clark, R., Hauth, D., Hill, F., Kroll, R., Luis, G., Mills, N., Purdy, T., Henney, C., Holland, D., Winter, J.: 2011, {\it Bull. Am. Astron. Soc.} {\bf 43}, 17.45. 
\bibitem[Howard \emph{et al.}(2008)]{2008SSRv..136...67H}Howard, R.A., Moses, J.D., Vourlidas, A., Newmark, J.S., Socker, D.G., Plunkett, S.P., Korendyke, C.M., Cook, J.W., Hurley, A., Davila, J.M., Thompson, W.T., St Cyr, O.C., Mentzell, E., Mehalick, K., Lemen, J.R., Wuelser, J.P., Duncan, D.W., Tarbell, T.D., Wolfson, C.J., Moore, A., Harrison, R.A., Waltham, N.R., Lang, J., Davis, C.J., Eyles, C.J., Mapson-Menard, H., Simnett, G.M., Halain, J.P., Defise, J.M., Mazy, E., Rochus, P., Mercier, R., Ravet, M.F., Delmotte, F., Auchere, F., Delaboudiniere, J.P., Bothmer, V., Deutsch, W., Wang, D., Rich, N., Cooper, S., Stephens, V., Maahs, G., Baugh, R., McMullin, D., Carter, T.: 2008, {\it Space Sci. Rev.} {\bf 136}, 67. 
\bibitem[Inhester(2006)]{2006astro.ph.12649I}Inhester, B.: 2006, \arxivurl{astro-ph/0612649}. 
\bibitem[Innes, McIntosh, and Pietarila(2010)]{2010A&A...517L...7I}Innes, D.E., McIntosh, S.W., Pietarila, A.: 2010, {\it Astron. Astrophys.} {\bf 517}, L7, \doiurl{10.1051/0004-6361/201014366}.
\bibitem[Joshi \emph{et al.}(2013)]{2013ApJ...771...65J} Joshi, N.C., Srivastava, A.K., Filippov, B., Uddin, W., Kashyap, P., Chandra, R.\ 2013, {\it Astrophys. J.} {\bf 771}, 65,\doiurl{10.1088/0004-637X/771/1/65}.
\bibitem[Joshi \emph{et al.}(2018)]{2018MNRAS.476.1286J}Joshi, N.C., Nishizuka, N., Filippov, B., Magara, T., Tlatov, A.G.: 2018, {\it Mont. Not. Roy. Astron. Soc.} {\bf 476}, 1286, \doiurl{10.1093/mnras/sty322}.
\bibitem[Kilpua, Koskinen, and Pulkkinen(2017)]{2017LRSP...14....5K}Kilpua, E., Koskinen, H.E.J., Pulkkinen, T.I.: 2017, {\it Liv. Rev. Solar Phys.} {\bf 14}, 5, \doiurl{10.1007/s41116-017-0009-6}.
\bibitem[Kusano \emph{et al.}(2004)]{2004ApJ...610..537K}Kusano, K., Maeshiro, T., Yokoyama, T., Sakurai, T.: 2004, {\it Astrophys. J.} {\bf 610}, 537, \doiurl{10.1086/421547}.
\bibitem[Leinweber, Russell, and Angelopoulos(2008)]{2008AGUFMSM11B1611L}Leinweber, H.K., Russell, C.T., Angelopoulos, V.: 2008, {\it AGU Fall Meeting Abs.}, SM11B-1611, \doiurl{10.1088/0957-0233/19/5/055104}.
\bibitem[Low(1977)]{1977ApJ...212..234L}Low, B.C.: 1977, {\it Astrophys. J.} {\bf 212}, 234, \doiurl{10.1086/155042}.
\bibitem[Lynch \emph{et al.}(2004)]{2004ApJ...617..589L}Lynch, B.J., Antiochos, S.K., MacNeice, P.J., Zurbuchen, T.H., Fisk, L.A.: 2004, {\it Astrophys. J.} {\bf 617}, 589, \doiurl{10.1086/424564}.
\bibitem[Lynch \emph{et al.}(2016)]{2016JGRA..12110677L}Lynch, B.J., Masson, S., Li, Y., DeVore, C.R., Luhmann, J.G., Antiochos, S.K., Fisher, G.H.: 2016, {\it J. Geophys. Res. A} {\bf 121}, 10, \doiurl{10.1002/2016JA023432.}.
\bibitem[Ma \emph{et al.}(2010)]{2010ApJ...722..289M}Ma, S., Attrill, G.D.R., Golub, L., Lin, J.: 2010, {\it Astrophys. J.} {\bf 722}, 289, \doiurl{10.1088/0004-637X/722/1/289.}.
\bibitem[Mishra and Srivastava(2019)]{2019ApJ...874...57M}Mishra, S.K., Srivastava, A.K.: 2019, {\it Astrophys. J.} {\bf 874}, 57, \doiurl{10.3847/1538-4357/ab06f2.  }.
\bibitem[Mishra \emph{et al.}(2018)]{2018ApJ...856...86M}Mishra, S.K., Singh, T., Kayshap, P., Srivastava, A.K.: 2018, {\it Astrophys. J.} {\bf 856}, 86, \doiurl{10.3847/1538-4357/aaae03}.
\bibitem[Moore and Labonte(1980)]{1980IAUS...91..207M}Moore, R.L., Labonte, B.J.: 1980, In: Dryer, M., Tandberg-Hanssen, E. (eds.) {\it Solar and Interplanetary Dynamics} IAU Symp. {\bf 91}, 207, Reidel, Dordrecht.
\bibitem[Moore, Sterling, and Falconer(2015)]{2015ApJ...806...11M}Moore, R.L., Sterling, A.C., Falconer, D.A.: 2015, {\it Astrophys. J.} {\bf 806}, 11, \doiurl{10.1088/0004-637X/806/1/11.  }.
\bibitem[Moore \emph{et al.}(2001)]{2001ApJ...552..833M}Moore, R.L., Sterling, A.C., Hudson, H.S., Lemen, J.R.: 2001, {\it Astrophys. J.} {\bf 552}, 833, \doiurl{10.1086/320559}.

\bibitem[Nieves-Chinchilla \emph{et al.}(2013)]{2013ApJ...779...55N}Nieves-Chinchilla, T., Vourlidas, A., Stenborg, G., Savani, N.P., Koval, A., Szabo, A., Jian, L.K.: 2013, {\it Astrophys. J.} {\bf 779}, 55, \doiurl{10.1088/0004-637X/779/1/55.}.
\bibitem[Nitta and Mulligan(2017)]{2017SoPh..292..125N}Nitta, N.V., Mulligan, T.: 2017, {\it Solar Phys.} {\bf 292}, 125, \doiurl{s11207-017-1147-7}.
\bibitem[Pesnell, Thompson, and Chamberlin(2012)]{2012SoPh..275....3P}Pesnell, W.D., Thompson, B.J., Chamberlin, P.C.: 2012, {\it Solar Phys.} {\bf 275}, 3, \doiurl{10.1007/s11207-011-9841-3}.
\bibitem[O'Kane \emph{et al.}(2019)]{2019ApJ...882...85O}O'Kane, J., Green, L., Long, D.M., Reid, H.: 2019, {\it Astrophys. J.} {\bf 882}, 85, \doiurl{10.3847/1538-4357/ab371b.}.
\bibitem[Pevtsov, Panasenco, and Martin(2012)]{2012SoPh..277..185P}Pevtsov, A.A., Panasenco, O., Martin, S.F.: 2012, {\it Solar Phys.} {\bf 277}, 185, \doiurl{10.1007/s11207-011-9881-8}. 
\bibitem[Richardson and Cane(2011)]{2011SpWea...9.7005R}Richardson, I.G., Cane, H.V.: 2011, {\it Space Weather} {\bf 9}, S07005, \doiurl{10.1029/2011SW000670}.
\bibitem[Robbrecht, Patsourakos, and Vourlidas(2009)]{2009ApJ...701..283R}Robbrecht, E., Patsourakos, S., Vourlidas, A.: 2009, {\it Astrophys. J.} {\bf 701}, 283, \doiurl{10.1088/0004-637X/701/1/283}.
\bibitem[Schmieder, D{\'e}moulin, and Aulanier(2013)]{2013AdSpR..51.1967S}Schmieder, B., D{\'e}moulin, P., Aulanier, G.: 2013, {\it Adv. Space Res.} {\bf 51}, 1967, \doiurl{10.1016/j.asr.2012.12.026}.
\bibitem[Srivastava \emph{et al.}(2010)]{2010ApJ...715..292S}Srivastava, A.K., Zaqarashvili, T.V., Kumar, P., Khodachenko, M.L.: 2010, {\it Astrophys. J.} {\bf 715}, 292, \doiurl{10.1088/0004-637x/715/1/292}.
\bibitem[Srivastava and Venkatakrishnan(2002)]{2002GeoRL..29.1287S}Srivastava, N., Venkatakrishnan, P.: 2002, {\it Geophys. Res. Lett.} {\bf 29}, 1287, \doiurl{10.1029/2001GL013597}.
\bibitem[Sterling and Moore(2004)]{2004ApJ...602.1024S}Sterling, A.C.~and Moore, R.L.: 2004, {\it Astrophys. J.} {\bf 602}, 1024, \doiurl{10.1086/379763}.
\bibitem[Sterling \emph{et al.}(2015)]{2015Natur.523..437S}Sterling, A.C., Moore, R.L., Falconer, D.A., Adams, M.: 2015, {\it Nature} {\bf 523}, 437, \doiurl{10.1038/nature14556.}.
\bibitem[van Ballegooijen and Martens(1989)]{1989ApJ...343..971V}van Ballegooijen, A.A., Martens, P.C.H.: 1989, {\it Astrophys. J.} {\bf 343}, 971, \doiurl{10.1086/167766}.
\bibitem[Vourlidas \emph{et al.}(2011)]{2011ApJ...733L..23V}Vourlidas, A., Colaninno, R., Nieves-Chinchilla, T., Stenborg, G.: 2011, {\it Astrophys. J.} {\bf 733}, L23, \doiurl{10.1088/2041-8205/733/2/L23}.
\bibitem[Webb(2015)]{2015ASSL..415..411W}Webb, D.F.: 2015, In: Vial, J.-C., Engvold, O. (eds.) Eruptive Prominences and Their Association with Coronal Mass Ejections, {\it Solar Prominences} \textit{Astrophys. Space Sci. Lib} {\bf 415}, Springer Cham, Heidelberg, 411, \doiurl{10.1007/978-3-319-10416-4}.
\bibitem[Webb and Howard(2012)]{2012LRSP....9....3W}Webb, D.F., Howard, T.A.: 2012, {\it Liv. Rev. Solar Phys.} {\bf 9}, 3, \doiurl{10.12942/lrsp-2012-3}.
\bibitem[Wuelser \emph{et al.}(2004)]{2004SPIE.5171..111W}Wuelser, J.-P., Lemen, J.R., Tarbell, T.D., Wolfson, C.J., Cannon, J.C., Carpenter, B.A., Duncan, D.W., Gradwohl, G.S., Meyer, S.B., Moore, A.S., Navarro, R.L., Pearson, J.D., Rossi, G.R., Springer, L.A., Howard, R.A., Moses, J.D., Newmark, J.S., Delaboudinière, J.-P., Artzner, G.E., Auchère, F., Bougnet, M., Bouyries, P., Bridou, F., Clotaire, J.-Y., Colas, G., Delmotte, F., Jerome, A., Lamare, M., Mercier, R., Mullot, M., Ravet, M.-F., Song, X., Bothmer, V., Deutsch, W.: 2004, {\it Telescopes and Instrumentation for Solar Astrophysics}, {\it Proc. SPIE}, {\bf 5171}, 111, \doiurl{10.1117/12.506877}.
\bibitem[Yang \emph{et al.}(2011)]{2011ApJ...732L...7Y}Yang, S., Zhang, J., Li, T., Liu, Y.: 2011, {\it Astrophys. J.} {\bf 732}, L7, \doiurl{10.1088/2041-8205/732/1/L7.}.
\bibitem[Yokoyama and Kamide(1997)]{1997JGR...10214215Y}Yokoyama, N., Kamide, Y.: 1997, {\it J. Geophys. Res.} {\bf 102}, 14215, \doiurl{10.1029/97JA00903}.
\bibitem[Yurchyshyn, Abramenko, Tripathi(2009)]{2009ApJ...705..426Y}Yurchyshyn, V., Abramenko, V., Tripathi, D.: 2009, {\it Astrophys. J.} {\bf 705}, 426, \doiurl{10.1088/0004-637X/705/1/426.}.
\bibitem[Zhang \emph{et al.}(2007)]{2007JGRA..11212103Z}Zhang, J., Richardson, I.G., Webb, D.F., Gopalswamy, N., Huttunen, E., Kasper, J., Nitta, N.V., Poomvises, W., Thompson, B.J., Wu, C.-C., Yashiro, S., Zhukov, A.N.: 2007, {\it J. Geophys. Res. A} {\bf 112}, A12103, \doiurl{10.1029/2007JA012891}.
\end{thebibliography}
\end{document}